\documentclass[acmsmall,screen, nonacm]{acmart}
\AtBeginDocument{%
  }

\setcopyright{acmlicensed}
\copyrightyear{2026}
\acmYear{2026}
\acmDOI{XXXXXXX.XXXXXXX}

\acmJournal{JACM}
\acmVolume{37}
\acmNumber{4}
\acmArticle{111}
\acmMonth{8}

\usepackage{soul}
\usepackage{tikz}
\usepackage{wrapfig}
\usepackage[table]{xcolor}
\definecolor{grayBlue}{HTML}{5F7893}
\usepackage{booktabs}
\usepackage{siunitx}
\usepackage{multirow}
\usepackage{tcolorbox}
\usepackage{enumitem}
\definecolor{quoteblue}{HTML}{1F5A78}
\newcommand{\pquote}[1]{{\color{quoteblue}``\textit{#1}''}}

\newcommand{\blogs}[1]{\unskip~{\scriptsize[#1]}}
\newcommand{\repos}[1]{\unskip~{\scriptsize[#1]}}

\usepackage{xcolor}
\definecolor{epicE}{HTML}{21618C}

\usepackage{xspace}

\definecolor{epicE}{HTML}{2C5E99}
\definecolor{epicP}{HTML}{1C7A68}
\definecolor{epicI}{HTML}{6B4BA1}
\definecolor{epicC}{HTML}{A35F14}

\DeclareRobustCommand{\EPIC}{%
  \mbox{\normalfont\sffamily\bfseries
    \textcolor{epicE}{E}\kern0.04em
    \textcolor{epicP}{P}\kern0.04em
    \textcolor{epicI}{I}\kern0.04em
    \textcolor{epicC}{C}}\xspace}
\begin{document}


\title{The \EPIC Framework for Spec-Driven Development}




\author{Bruno Claudino}
\email{claudinomab@vcu.edu}
\authornotemark[1]
\affiliation{%
  \institution{Virginia Commonwealth University}
  \country{USA}
}

\author{Ruizhe Xing}
\affiliation{%
  \institution{Oregon State University}
  \country{USA}}
\email{xingru@oregonstate.edu}

\author{Yuanhao Zuo}
\affiliation{%
  \institution{Virginia Commonwealth University}
  \country{USA}
}
\email{zuoy@vcu.edu}

\author{Tyler Menezes}
\affiliation{%
 \institution{CodeDay}
 \country{USA}
 }
 \email{tylermenezes@codeday.org}

\author{Anita Sarma}
\affiliation{%
  \institution{Oregon State University}
  \country{USA}
}
\email{anita.sarma@oregonstate.edu}

\author{Kostadin Damevski}
\affiliation{%
  \institution{Virginia Commonwealth University}
  \country{USA}}
\email{kdamevski@vcu.edu}

\author{Zixuan Feng}
\affiliation{%
  \institution{Virginia Commonwealth University}
  \country{USA}}
\email{fengz3@vcu.edu}


\renewcommand{\shortauthors}{Trovato et al.}

\begin{abstract}

One practitioner we interviewed said their team writes ``must'' instead of ``should'' when instructing a coding agent, because the agent may treat ``should'' as optional. 
Small wording choices matter because agents often fill gaps in their instructions with their own assumptions. Spec-driven development (SDD) asks developers to write a specification, plan, and tasks before the agent writes code. SDD frameworks provide templates for these artifacts, but the templates do not help developers judge whether they have written enough or clearly enough. We studied what good SDD specifications contain. We scored the artifacts of 114 open-source SDD repositories against ISO/IEC/IEEE~29148 and derived practices from the highest-scoring ones. 
The resulting framework, \EPIC, has 40 practices in 10 quality dimensions that guide developers in making expectations and decisions explicit in specifications, plans, and tasks for coding agents.
The majority of SDD practitioners (N=15) endorsed every practice. Repositories in the top third of specification quality spent 11.8\% of their commits on bug fixes, compared with 20.4\% in the bottom third, and had a median of 4x as many contributors. Developers can use \EPIC to find and fill the gaps in a specification before the agent does.
\end{abstract}

\ccsdesc[300]{Human-centered computing}

\received{05 October 2026}
\received[revised]{12 March 2009}
\received[accepted]{5 June 2009}

\maketitle

\section{Introduction}
\label{sec:introduction}

Ask a colleague to make a query ``fast,'' and they will likely ask how fast, under what workload, and how the result will be checked. Ask an AI coding agent, and it may fill the gap with its own assumptions and return code that appears to answer the request~\cite{mu2024clarifygpt}. Agents struggle to recognize when information is missing, even when asking would improve their solutions~\cite{vijayvargiya2026ambig}. As one practitioner we interviewed put it, the agent is \pquote{making a decision even if I don't} (P10). Once the code exists, it is difficult to tell which assumptions came from the developer and which came from the agent.

The underlying problem is not new. Software has always been only as good as the account of what it should do, and requirements engineering has long held that this account must state not only the required behavior but also the assumptions, constraints, and operating conditions under which the software must work~\cite{heninger1980specifying,zave1997darkcorners}. Incomplete and inconsistent requirements remain common problems in industry, associated with rework and overruns~\cite{mendez2017namingthepain}. Coding agents raise the stakes. Producing code faster does not necessarily mean producing better software: in a study of Cursor adoption, gains in development velocity were temporary, while increases in static-analysis warnings and code complexity persisted~\cite{he2026speed}. As developers increasingly orchestrate agents rather than write every line themselves, the specification becomes the main place where they decide what gets built, and the one artifact they can review before any code exists.

\emph{Spec-driven development} (SDD) puts this specification at the center of the workflow: a structured natural-language specification drives the agent's implementation~\cite{agarwal2026specmine}. The practice barely existed before 2025 and now spans more than 70,000 public GitHub repositories~\cite{agarwal2026specmine}. Yet developers have little basis for judging whether a specification says enough. SDD frameworks such as GitHub Spec Kit, Kiro, and OpenSpec provide templates and workflow steps~\cite{github2025speckit,kiro2025requirementsfirst,fissionopenspec}, but a template shows where information belongs, not what is sufficient, and each framework structures it differently. A specification can fill every section and still leave assumptions unstated or contradict its plan. Requirements-quality standards define good requirements, individually and as a set~\cite{ISO/IEC/IEEE-29148-2018,femmer2017rapid}, but not the plans and tasks that SDD adds and that must agree with them. CAFE(S) defines qualities of the context an agent receives as a shared language rather than a measurement system~\cite{houck2026cafes}. Developers therefore lack concrete, empirically grounded guidance on what to articulate in SDD artifacts, and little evidence links requirements quality to development outcomes~\cite{frattini2023requirements}.

To our knowledge, this is the first study to derive SDD specification practices from both requirements-engineering standards and the artifacts developers write. We investigate two research questions:

\begin{itemize}[leftmargin=*, labelindent=0pt, nosep, itemsep=1pt]
    \item \textbf{RQ1:} What specification practices are used in SDD, and how do they address specification quality?
    \item \textbf{RQ2:} How does specification quality relate to software outcomes, and how do practitioners assess the clarity and usefulness of the identified practices?
\end{itemize}

Through a mixed-methods study, we established a quality lens from ISO/IEC/IEEE~29148:2018 and SDD framework guidance, applied it to artifacts from 114 open-source SDD repositories, derived practices from exemplar artifacts, and refined them using 30 grey-literature sources. The result, \emph{\EPIC}, organizes 40 practices for writing and reviewing SDD artifacts into ten quality dimensions and four categories: \textbf{E}xpressive, \textbf{P}recise, \textbf{I}nterpretable, and \textbf{C}ompartmentalized. Our study yields three main findings:
\begin{enumerate}[leftmargin=*, labelindent=0pt, nosep]
    \item \textbf{SDD shifts the unit of quality from the single requirement to the artifact set.} The three new dimensions \EPIC adds to the seven grounded in ISO/IEC/IEEE~29148:2018 concern a specification, a plan, and tasks that must agree with one another. They cover the dependencies and constraints requirements rely on, separating required behavior from implementation decisions, and recording planning information.
    \item \textbf{Practitioners agree on the practices and on when they pay off.} None of the 30 grey-literature sources added a practice, and a majority of the 15 practitioners we interviewed endorsed every one of the 40. Their explanations show when a practice matters, so teams can adopt practices as a project grows rather than all at once.
    \item \textbf{Specification quality tracks with how repositories evolve.} Repositories in the top third of scores spent 11.8\% of their commits on bug fixes versus 20.4\% in the bottom third, had 15\% lower median code complexity, and had a median of 12.5 contributors versus 3. We treat these observational associations as hypothesis-generating; bug and vulnerability density results were mixed.
\end{enumerate}

In summary, this paper contributes: (1)~\EPIC, an empirically grounded framework of 40 specification practices for SDD artifacts; (2)~practitioner accounts of when these practices are worth their cost; and (3)~exploratory evidence that specification quality is associated with code quality and development activity. These contributions give developers concrete criteria for reviewing a specification before any code exists, an agent skill that applies them, and a structure for benchmarking how agents act on specifications.

\vspace{-3mm}

\section{Background}
\label{sec:background}

Software engineering has spent more than half a century trying to write down what a program should do before writing the program \cite{hoare1969axiomatic,heninger1980specifying,zave1997darkcorners}.

\textbf{Fifty Years of Trying to Capture Intent.}
Early work on software specification and requirements engineering emphasized making intended behavior, interfaces, and assumptions explicit, helping engineers understand what to build and assess whether an implementation meets those expectations~\cite{hamilton1976higherorder,heninger1980specifying,hoare1969axiomatic,zave1997darkcorners,meyer1992designbycontract}. These approaches address a shared problem: when intent and implementation are separated, they can drift apart. Industrial surveys report incomplete and inconsistent requirements as common problems leading to rework and budget or schedule overruns~\cite{mendez2017namingthepain,wagner2019statusquo}. Approaches such as Test-Driven Development and Behavior-Driven Development were later developed to make intent more visible and easier to check~\cite{beck2002tdd,north2006bdd}. These approaches developed primarily around the assumption that a human would write the code.

\textbf{From Vibe Coding to Specification-Driven Development.}
AI code generation allows natural-language instructions to guide implementation~\cite{chen2021codex}. In \emph{vibe coding}, developers describe desired outcomes and refine prompts after running the generated code~\cite{chou2026vibecoding,fawzy2026vibecoding}. As conversations grow, requirements can become implicit or scattered across prompts, making completeness and consistency difficult to assess~\cite{ullrich2025requirementstocode,huang2026regressionaccumulation, baumann2026swe, yang2026first}. Developers often struggle to understand, debug, and steer AI-generated code~\cite{vaithilingam2022expectation,liang2024usability}. SWE-chat documents user corrections in developer--agent interactions~\cite{baumann2026swe}, while repository research shows agents overlooking documented contribution rules~\cite{yang2026first}. These findings highlight that the difficulty is not simply whether the model can write code. It is whether the developer can make their intent clear enough for the model to act on it.



SDD has regained attention among researchers and practitioners working on AI-assisted software development~\cite{feng2025charting,piskala2026spec,diaz2026sddagentic}. In SDD, developers express the desired outcome in reviewable natural language that guides planning, implementation, and validation across sessions~\cite{diaz2026sddagentic,agarwal2026specmine}. These specifications serve as a shared reference point between humans and agents~\cite{diaz2026sddagentic,piskala2026spec,tang2026programmingbychat}. In practice, SDD takes the form of lightweight document pipelines. For example, Kiro separates requirements, design, and tasks; GitHub Spec Kit follows \texttt{Specify} $\rightarrow$ \texttt{Plan} $\rightarrow$ \texttt{Tasks} $\rightarrow$ \texttt{Implement}; and OpenSpec organizes proposals, specifications, designs, tasks, and archives \cite{kiro2025requirementsfirst,github2025speckit,fissionopenspec}. Despite their differences, these systems share a premise: natural-language intent should become a persistent artifact that mediates collaboration between humans and agents.

\textbf{The Unresolved Problem of Specification Quality.} A compiler may reject invalid input, and a colleague may ask for clarification. An agent may silently resolve ambiguity and return code that appears to meet the request~\cite{mu2024clarifygpt,vijayvargiya2026ambig}. Small differences in how a requirement is phrased can substantially alter model outputs \cite{mizrahi2024multiprompt}; issuing the same prompt twice can produce materially different implementations~\cite{ouyang2025nondeterminism}; and underspecified requirements are a measurable source of errors in generated code~\cite{mu2024clarifygpt}.

Prior work addresses parts of this problem. \citet{lucassen2016qus} developed a framework for evaluating user stories according to syntactic, semantic, and pragmatic criteria. Research on BDD shows that executable specifications can still suffer from duplication and weaknesses in readability, reuse, abstraction, and maintainability~\cite{binamungu2018maintaining,binamungu2020bddquality}. More recently, \citet{houck2026cafes} proposed CAFE(S), a conceptual framework identifying clarity, actionability, fidelity, efficiency, and security as desirable properties of agent context. 


These contributions raise an urgent question for SDD in the age of AI:  \emph{Which qualities and practices characterize specification and planning artifacts intended to guide AI-assisted implementation?} To address this need, we introduce \EPIC, an empirically grounded framework of 40 practices in ten quality dimensions for deciding what to tell coding agents and how to express it across specifications, plans, and tasks.


\section{Methodology}
\label{sec:methodology}

\begin{figure}[t]
    \centering
    \includegraphics[width=0.8\linewidth]{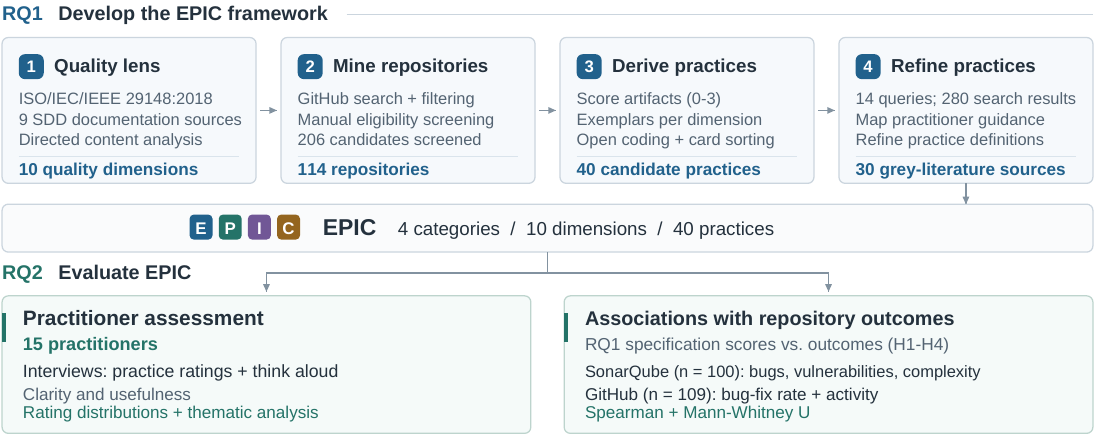}
  \caption{Overview of the study design.}
    \label{fig:method-overview}
    \vspace{-6mm}
\end{figure}

We used a mixed-methods design~\cite{storey2025guiding} to develop and evaluate \EPIC. We first identified and organized practices for addressing the quality of specification and planning artifacts in SDD (RQ1). We then evaluated \EPIC through practitioners' views of the framework's clarity and usefulness and examined associations between specification quality and indicators of code quality and development activity (RQ2). Figure~\ref{fig:method-overview} summarizes the study design.

\subsection{RQ1. Developing the \EPIC Framework}

We developed \EPIC in four steps. (1) We drew on ISO/IEC/IEEE~29148:2018~\cite{ISO/IEC/IEEE-29148-2018} and SDD tool guidance \cite{kiro2025introducing, github2025speckit} to establish ten specification quality dimensions, which formed an initial \textit{quality lens}. (2) We collected specification and planning artifacts from open-source SDD repositories. (3) We assessed these artifacts against the quality lens and analyzed exemplars for each dimension to identify concrete practices for addressing it. This analysis yielded 40 candidate practices. (4) We examined practitioner guidance through a grey literature review~\cite{garousi2019guidelines} to examine their support for the practices, refine their definitions, and identify concerns not captured by the repository analysis.

\subsubsection{Establishing a Specification Quality Lens.}

Requirements engineering standards define characteristics used in empirical research and quality assurance to assess requirements quality~\cite{montgomery2022empirical, femmer2017rapid}. Prior studies have also drawn on ISO-defined characteristics to examine requirements and software quality, including in AI-assisted development~\cite{paiva2026issue,feng2025domains}. Following this approach, we selected ISO/IEC/IEEE~29148:2018, a requirements engineering standard that defines the characteristics of good software requirements. 


\textbf{Incorporating SDD tool guidance.}
To interpret these characteristics in the context of SDD, we complemented them with documentation and specification templates from popular SDD kits (N=9) (i.e., GitHub Spec Kit \cite{github2025speckit}, Kiro~\cite{kiro2025introducing}). We used these documents to understand how requirements inform design decisions and implementation tasks in SDD, and what information agents need at each stage.

\textbf{Coding and reconciling quality expectations.}
We conducted a directed qualitative content analysis~\cite{hsieh2005three}, using the standard's requirements-quality characteristics and definitions as the initial codebook. Four researchers independently applied this codebook to the SDD kits (N=9) documentation, treating each passage expressing a quality expectation or specification practice as a coding unit. They applied existing codes to matching passages and assigned provisional codes to additional quality concerns in SDD. In weekly meetings, they compared their coding and reconciled differences through negotiated agreement~\cite{campbell2013coding}.

\textbf{Defining the ten quality dimensions.}
This analysis yielded ten quality dimensions for specification and planning artifacts in SDD. Seven retained characteristics are named in the ISO/IEC/IEEE~29148:2018: \textit{Unambiguous}, \textit{Singular}, \textit{Verifiable}, \textit{Conforming}, \textit{Complete}, \textit{Consistent}, and \textit{Comprehensible}~\cite{ISO/IEC/IEEE-29148-2018}. The remaining three adapt or synthesize the standard and tool guidance for SDD. \textit{Dependency Control} adapts \textit{Feasible} to focus on the documented assumptions, dependencies, and constraints that condition whether requirements can be realized~\cite{ISO/IEC/IEEE-29148-2018,github2025speckit}. \textit{Scope Discipline} combines guidance on requirements abstraction and the separation of requirements from implementation decisions~\cite{ISO/IEC/IEEE-29148-2018,github2025speckit}. \textit{Implementation-Planning Information} captures the planning and task information emphasized by the SDD tools~\cite{github2025speckit,kiro2025introducing}, informed by the standard's provisions for assumptions, dependencies, and requirements allocation~\cite{ISO/IEC/IEEE-29148-2018}. See the definitions of the ten quality dimensions in \cite{epic_website_2026}.

\vspace{-3mm}

\subsubsection{Mining OSS Repositories for SDD Artifacts}
The quality dimensions describe what a good specification achieves, but not how a specification achieves it. To identify practices for addressing those properties, we needed to understand how SDD projects documented requirements and implementation guidance. We planned to use the ten quality dimensions to assess these artifacts and select examples from which to derive practices.

\textbf{Discovering candidate repositories.}
We first collected specification and planning artifacts from GitHub. Because SDD workflows vary in their tools, filenames, and directory structures~\cite{agentreadmes2025,agentsmdsmells2026}, we used two types of discovery signals to identify candidate repositories: (1) specification-related filenames and framework-specific structures. Filenames such as \texttt{spec.md}, \texttt{requirement.md}, and \texttt{PRD.md} helped identify candidates that might use SDD without a dedicated framework. (2) We also searched for structures associated with Spec Kit and OpenSpec. For Spec Kit, these included the \texttt{.specify/} directory and artifacts such as \texttt{spec.md}, \texttt{plan.md}, \texttt{constitution.md}, and task files. For OpenSpec, they included \texttt{openspec/changes} and artifacts such as \texttt{proposal.md}, \texttt{design.md}, and specification files. We executed the searches through the GitHub Code Search API on April 25, 2026. This step yielded 850 candidate repositories.


\textbf{Filtering and screening candidate repositories.}
We narrowed the candidate set through repository-level filtering and manual eligibility screening~\cite{miller2022did,baltes2022sampling}.
We applied three filters adapted from prior studies. First, we removed forks and duplicate repositories to reduce duplication in the sample~\cite{kalliamvakou2016promises,munaiah2017curating, feng2022case}. Second, we retained repositories whose relevant specification artifacts were written in English~\cite{afroz2026ai}. Third, we required at least ten GitHub stars as a minimum signal of community interest inspired by existing studies~\cite{miller2022did,agentreadmes2025}. This step yielded 206 repositories.

We then screened the remaining repositories for evidence of SDD use.
Before screening, four authors defined the inclusion criterion: a repository was eligible if it contained specification artifacts intended to guide or constrain an AI agent's implementation.  Two authors independently screened the remaining repositories by inspecting their documentation and artifacts, then resolved disagreements through negotiated agreement~\cite{garrison2006negotiated}. Of the 206 repositories screened, 114 met the criterion and formed the final dataset.

\subsubsection{Deriving Candidate Specification Practices from SDD Artifacts}


Using the SDD repository dataset, we examined how specification and planning artifacts addressed the ten quality dimensions in practice. For each dimension, we selected artifacts with strong evidence of that quality and analyzed them to identify the practices they used. The analysis had five steps:

\emph{\textcircled{1} Operationalizing the quality dimensions.}
To assess repositories, we developed a four-point rubric based on the definitions and descriptions of the ten quality dimensions. We scored each repository on each dimension using evidence from its specification and planning artifacts: 0 indicated little or no evidence of the quality, 1 partial evidence, 2 mostly consistent evidence, and 3 consistent evidence throughout the artifacts. For example, for \textit{Singular} (E2), a repository scored 0 when requirements frequently bundled multiple capabilities or concerns into one statement; 1 when some requirements were singular but compound statements remained common; 2 when most requirements expressed a single capability, characteristic, constraint, or quality, with occasional compound statements; and 3 when all requirements did so consistently. The full rubric is available in the replication package~\cite{suppl}.

\emph{\textcircled{2} Calibrating the application of the rubric.}
Two researchers then independently assessed the same repositories in two calibration rounds and discussed every disagreement after each round. We first calibrated the rubric on approximately 10\% of the dataset (13 repositories). After discussing scoring differences, we conducted a second round on approximately another 20\% (23 repositories) to check whether the researchers could apply the rubric consistently to new repositories. These rounds yielded 130 and 230 paired dimension ratings.

We assessed agreement using exact agreement, quadratic-weighted Cohen's $\kappa_w$, and ordinal Krippendorff's $\alpha$ \cite{cohen1968weighted}. The latter two measures account for chance agreement and distinguish smaller from larger disagreements on the ordinal scale. We interpreted $\kappa_w$ using Landis and Koch's categories~\cite{landis1977measurement}. In the first round, the researchers agreed on 104 of 130 ratings (80.0\%), and no disagreement exceeded one scale point. Weighted agreement was almost perfect ($\kappa_w = 0.82$; $\alpha = 0.94$). In the second round, exact agreement was 83\%, with substantial weighted agreement ($\kappa_w = 0.78$; $\alpha = 0.80$). These results supported proceeding with the division and scoring of the remaining repositories.

\emph{\textcircled{3} Scoring the remaining repositories independently.}
The two researchers divided the remaining 78 repositories between them. Each researcher scored their assigned repositories and discussed difficult or ambiguous cases with the other researcher.

\emph{\textcircled{4} Selecting exemplars for each dimension.}
During the analysis, we observed a repository could be an exemplar for \textit{Unambiguous} without being one for \textit{Complete}; we therefore selected exemplars separately for each dimension.  For each dimension, we selected the top-scoring 20\% of repositories and used their artifacts as examples for the subsequent analysis~\cite{baltes2022sampling}. When repositories shared the cutoff score, we included all of them. The step yielded between 28 and 37 exemplar repositories per dimension.

\emph{\textcircled{5} Deriving practices from the exemplars.}
For each dimension, two researchers independently coded selected exemplars to identify the specification practices they contained. Using open coding with constant comparison~\cite{saldana2021coding}, they compared new passages with those already coded, reusing a code when it described the same practice and creating a new code when no existing code fit. They met weekly to compare their coding, merge duplicate codes, and resolve disagreements through negotiated agreement~\cite{garrison2006negotiated}. Across the ten dimensions, they coded 5,675 segments.

The researchers then used card sorting to group codes describing similar ways of addressing each dimension~\cite{zimmermann2016card}. They clarified the boundaries between groups and named and defined each group as a specification practice under that dimension. This process yielded 40 practices, each describing a recurring way the exemplar artifacts addressed a quality dimension.

\subsubsection{Refining Practices with Practitioner Guidance from Grey Literature}
To validate and refine the practices derived from repositories, we reviewed practitioner guidance in blogs, technical guides, framework documentation, and repositories~\cite{agentreadmes2025,sdcgen2026, gao2026registry, afroz2026ai}. We examined whether these sources recommended the practices, identified missing concerns, or suggested revisions, drawing on their explanations of why and how the practices should be applied. We followed guidelines for grey literature research (GLR) in software engineering~\cite{garousi2019guidelines}.

\emph{\textcircled{1} Searching for sources.}
To identify relevant grey literature, we drew on terminology from the Spec Kit and Kiro documentation~\cite{github2025speckit,kiro2025introducing} to formulate 14 search queries, organized into three groups. The first targeted SDD terminology, including ``spec-driven development'' and ``specification-driven development,'' alone or combined with ``LLM.'' The second targeted frameworks and artifacts, including Spec Kit, OpenSpec, \texttt{AGENTS.md}, \texttt{CLAUDE.md}, and \texttt{copilot-instructions.md}. The third restricted searches to GitHub to locate repository-based guidance and discussions. We collected the first 20 Google results for each query, following an approach also used in prior software engineering GLR~\cite{sotolani2025exploring,fawzy2026vibecoding}. The step yielded 280 artifacts.

\emph{\textcircled{2} Screening sources.} 
Before screening, we defined three inclusion and five exclusion criteria. A source had to meet all three inclusion criteria: (1) address writing or using specifications for AI-assisted development (\textit{topic relevance}); (2) recommend concrete practices or provide examples (\textit{actionable content}); and (3) make its full text freely available (\textit{accessibility}). We excluded sources that (1) used ``specification'' in a sense unrelated to guiding AI agents, such as test, hardware, or API specifications; (2) mentioned SDD or AI coding tools only in passing, or consisted solely of news, announcements, marketing, or installation instructions; (3) duplicated or republished a source already retrieved; (4) were inaccessible (broken, paywalled, or login-only) or non-textual without a transcript; or (5) were peer-reviewed academic papers, which fall outside grey literature. This process yielded 30 artifacts.

\emph{\textcircled{3} Mapping guidance to practices.}
We then conducted a directed qualitative content analysis using the 40 practices as the initial codebook~\cite{hsieh2005three}. Two researchers independently coded artifacts focusing on specification-writing practices, recommendations, or examples from each source, mapping each passage to the practices it supported. They then compared their coding and resolved disagreements through negotiated agreement~\cite{garrison2006negotiated}. The review did not lead to additional practices or dimensions. Each concern identified in the 30 sources either mapped to an existing practice or fell outside the study's scope. The sources provide explanations of why practices matter and how to apply them. We used these explanations to revise the definitions of practices, clarifying what each practice requires and how a specification author can address it (see the GLR mapping in \cite{epic_website_2026}).

\subsubsection{Assembling the \EPIC Framework}
We next organized the dimensions and practices into a framework. Four researchers compared each dimension's definition and associated practices, then grouped the dimensions by the questions they address about a specification. They resolved disagreements through negotiated agreement~\cite{garrison2006negotiated}. The resulting \textbf{\EPIC framework} (Figure~\ref{fig:results_evaluation}) comprises four categories, ten quality dimensions, and 40 specification practices.


\vspace{-2mm}

\subsection{RQ2. Evaluating the EPIC Framework}


\subsubsection{Practitioner Evaluation of Clarity and Usefulness}

We first assessed how practitioners interpreted \EPIC's practices, whether they considered them useful for constructing specifications, and what they would change. Inspired by the established approaches to taxonomy evaluation~\cite{usman2017taxonomies,nickerson2013taxonomy}, we conducted individual semi-structured interviews~\cite{feng2022case, feng2025domains}.

\emph{Designing and piloting the interviews.}
Each interview lasted approximately 60-90 minutes and began with a brief demographic questionnaire. We then asked participants about their use of SDD, including their development practices, motivations, challenges, and learning experiences. These questions provided context for interpreting their subsequent assessments.

Participants were then asked to review the framework's 10 dimensions and 40 metrics through a custom web application~\cite{epic_website_2026}. Each metric appeared on a card with an explanation and a real example from GitHub exemplars. For each metric, participants rated their agreement with the statement on a five-point Likert scale ranging from strongly disagree to strongly agree.

Throughout the interview, participants were asked to think aloud, explaining their interpretation of each metric and the reasoning behind their rating. Those unfamiliar with thinking aloud completed a brief practice exercise. We used follow-up questions to explore unclear wording, practical usefulness, overlap between metrics, and missing practices. Participants could suggest modifications and share examples from specifications they had written or reviewed. The interview concluded with questions about the framework's overall structure and how well it reflected SDD practice.

To encourage feedback and reduce acquiescence bias~\cite{heal1995response}, we emphasized that we were evaluating the framework, and explicitly welcomed disagreement and criticism. Dimensions appeared in randomized order to distribute potential ordering and fatigue effects across the framework's dimensions \cite{galesic2009effects}. Interviews were audio-recorded and transcribed with participants' consent. 

Before conducting the study, we performed three pilot interviews to assess the interview protocol and evaluation instrument, including the clarity of the instructions and the time required to review the framework. Based on the pilot feedback, we revised the interview flow and rephrased instructions and questions to make the interview easier to follow.

\emph{Participant recruitment and data analysis.}
We recruited software professionals from our professional network and used snowball sampling to recruit additional participants through their referrals. To ensure that participants have SDD experience, we first emailed potential participants to ask whether they had used SDD. We then invited only those who confirmed such experience to take part in the study. The study was approved by our university's IRB, and participants were offered a \$50 gift card. In total, we recruited 15 participants: seven backend or full-stack developers, two data/ML engineers, a QA engineer, a system architect, a project manager, an electrical engineer, and two software researchers. Fourteen had at least six years of development experience, eleven used specifications or structured context to guide AI coding tools often or very often, and twelve had used AI tools with written specifications for six months or more (see demographics in \cite{suppl}).


To understand why participants rated each practice as they did, we analyzed their explanations using inductive coding ~\cite{braun2006using} and card sorting~\cite{zimmermann2016card}. We organized the transcripts by practice, keeping each explanation linked to its corresponding rating. Two researchers independently developed codes from participants' explanations, then compared their coding and resolved disagreements through negotiated agreement~\cite{garrison2006negotiated}. They used card sorting to group related codes into themes capturing participants' reasons for agreement and disagreement~\cite{zimmermann2016card}. After analyzing the first 13 interviews, the results were approaching saturation and we stopped recruitment. We then completed and analyzed two additional interviews that had already been scheduled to confirm this assessment. Neither yielded new substantive feedback on the framework, satisfying our saturation criterion of two consecutive interviews without new findings. 

\vspace{-3mm}

\subsubsection{Specification Quality and Software Outcomes}

We also examined whether the specification qualities underlying \EPIC were associated with code quality and development activity. We focused on code quality and development activity because they address two fundamental concerns in software engineering: building dependable software and sustaining its maintenance and evolution~\cite{forsgren2021space,alami2024free}. Code quality provides evidence about the implementation, while development activity reflects the corrective work and contributor participation through which software evolves.

To conduct the analysis, we returned to the 114 GitHub repositories identified in RQ1 as using specifications to guide AI agents' implementation work. Each repository had received a score from 0 to 3 on each of the ten quality dimensions, based on evidence in its specification and planning artifacts. We summed these scores to obtain an overall specification-quality score between 0 and 30.

During the analysis, we had to disregard 5 repositories. 2 of them \repos{R60, R599} were deleted or moved to a different location by the time we performed the analysis. We also faced cloning problems with 3 others \repos{R63, R156, R236}, which did not allow us to perform the analysis. Hence, we perform the SonarQube and GitHub analysis in 109 repositories.

\emph{(1) Examining code quality: reliability, security, and structural maintainability.}
In a study of Cursor adoption, He et al.\ found temporary gains in development velocity but persistent increases in static-analysis warnings and code complexity~\cite{he2026speed}. We therefore examined whether specification quality was associated with reliability, security, and structural maintainability. Explicit descriptions of expected behavior and edge cases can improve the functional correctness of generated code~\cite{zi2025more}. Stating security expectations may similarly help agents avoid unsafe assumptions, addressing concerns raised by vulnerabilities in Copilot-generated code~\cite{pearce2022}. Clear scope and implementation guidance may also reduce ad hoc decisions and unnecessary branching, particularly when agents fail to recognize underspecified instructions~\cite{vijayvargiya2026ambig}. We therefore hypothesize that \emph{higher specification quality is associated with lower bug density (H1), lower vulnerability density (H2), and lower cyclomatic complexity per KLOC (H3).}

We attempted SonarQube analysis on 109 repositories. However, 9 repositories faced SonarQube limitations that resulted in incorrect or incomplete information: 4/9 could not be measured by SonarQube (2 because the analysis failed and 2 because they contained source code that was not analyzable) and 5/9 had fewer than 1,000 NCLOC. For that reason, for H1--H3 analysis, we used 100 repositories.  The analysis was conducted through a local SonarQube Community Build server (version 26.9) running in Multi-Quality Rule (MQR) mode, using SonarScanner CLI 8.0.1, the server's bundled analyzers (e.g., Python 5.31, JavaScript/TypeScript 13.8, Java 8.41, and Go 1.43), and the built-in \emph{Sonar way} quality profile for each language. For each repository, we made a clone of its default branch and analyzed the latest commit; analyses ran on September 12th, 2026.

\emph{(2) Examining corrective work and development activity.}
Beyond code quality, software depends on continued contributions to address problems, review changes, and develop new capabilities~\cite{alami2024free}. Specifications may support this work by giving contributors a shared understanding of what to build and how their tasks relate. Explicit requirements may also reduce corrective work caused by misunderstandings or omissions. We therefore hypothesize that \emph{higher specification quality is associated with a lower bug-fix commit rate and higher development activity (H4).}

To test H4, we collected repository metrics for 109 repositories through the GitHub REST API, including counts of pull requests, issues, contributors, merged pull requests, and commits. To examine corrective work, we used the bug-fix commit rate, motivated by prior research on corrective maintenance~\cite{amit2021corrective}. We calculated this rate as the share of scanned commits whose messages began with a fix pattern, such as \texttt{fix:}, \texttt{bugfix}, or \texttt{hotfix}, restricting the calculation to repositories with at least 20 scanned commits. The median repository had 136 pull requests, 50 issues, six contributors, and 771 commits on its default branch. Among repositories with at least 20 commits, the median share of bug-fix commits was 14.9\% (see~\cite{suppl} for details of each metric analyzed).

\emph{(3) Testing the hypotheses.} 
We used two complementary analyses to test H1--H4. First, we computed Spearman's rank correlation between the overall specification-quality score and each outcome measure across the full score range. Second, we compared outcomes between repositories with higher and lower specification quality. We divided the observed score distribution into tertiles, each intended to contain approximately one-third of the repositories, and compared the outer bands~\cite{gelman2009splitting, destefanis2026audit}. Such comparisons between upper and lower score tertiles have  been used in empirical software engineering~\cite{destefanis2026audit}. The observed boundaries yielded bottom ($\leq 18$), middle (19--22), and top ($\geq 23$) bands, with at least five score points separating the top and bottom bands. We kept repositories tied at each cutoff together in the corresponding outer band, so group sizes could differ. We compared outcomes between the top and bottom bands using the Mann--Whitney $U$ test and reported Cliff's $\delta$ as the effect size. 
\section{Results}

In this section, we first present the \EPIC framework for characterizing the quality of specification and planning artifacts. We then report practitioners' assessments of the practices and the associations between specification quality and repository outcomes.

\vspace{-3mm}

\subsection{The \EPIC Framework for Specification-Driven Development (SDD)}

\textit{\EPIC} is a multidimensional framework designed to support the development of high-quality software with SDD. It provides practices for writing and reviewing the specification and planning artifacts that guide AI agents' implementation work. The framework, as shown in Figure \ref{fig:results_evaluation}, organizes ten quality dimensions and 40 specification practices into four complementary categories: \textbf{E}xpressive, \textbf{P}recise, \textbf{I}nterpretable, and \textbf{C}ompartmentalized. An interactive version is available at \cite{epic_website_2026}.

\textbf{E: Expressive.} An AI agent needs to know what is required and how its work will be checked. An expressive specification communicates this through requirements with a clear intended meaning (\textit{Unambiguous}, E1), one identifiable obligation (\textit{Singular}, E2), and explicit criteria for checking whether that obligation has been met (\textit{Verifiable}, E3)~\cite{ISO/IEC/IEEE-29148-2018}.

\textit{E1. Unambiguity} 
An unambiguous specification makes the intended meaning of each requirement clear and identifies where that meaning remains unresolved. Markers such as \emph{[NEEDS CLARIFICATION]} identify unresolved decisions and show where further input is needed (E1.1)~\blogs{B4, B26}.\footnote{The prefix B identifies grey-literature sources listed in the supplementary material~\cite{epic_website_2026}.}
For settled decisions, explicit, measurable bounds state the expected behavior and the conditions under which it applies (E1.2). For example, one repository specified that ``\textit{Operators can query any SuzieQ-supported network table and receive results within 5 seconds for networks up to 500 devices}'' (R441).\footnote{Repository identifiers refer to entries in the \textit{Repositories} page; see~\cite{epic_website_2026}.}
The five-second limit states the required response time, while the 500-device limit bounds the applicability. The same repository listed the supported network tables, making clear which tables the requirement covered (E1.3)~\blogs{B27, B30}.
Normative wording makes clear whether a statement requires, permits, or prohibits an action. For example, ``\textit{It MUST NOT touch the persistent \texttt{MultiSweepService}/scheduled sweep config}'' (R573) states a binding prohibition. Consistent use of these terms and statement structures supports interpretation across requirements (E1.4). Explicit exclusions and verification boundaries further specify what each requirement and its checks cover (E1.5)~\blogs{B9, B10}.

\textit{E2. Singularity.} A singular requirement states one capability, property, or constraint. Details that define the same capability can stay together. For example, one repository specified ``\textit{R7. Support max\_depth parameter (default 8, max 15) to limit tree depth}'' (R80). The default and maximum values define the depth limit, so they belong in the same requirement (E2.1). The repository recorded a scope exclusion separately: ``\textit{This feature does NOT capture visual layout or screenshots}.'' Each statement can then be reviewed on its own. The other practices apply this approach to design facts, risks and assumptions, definitions and scenarios, and security properties (E2.2--E2.5) \blogs{B1, B29}.

\textit{E3. Verifiability.} A verifiable specification states what counts as satisfying a requirement, how satisfaction can be checked, what scenarios the checks must cover, and when verification is considered complete. For example, one repository documented ``\textit{Backend coverage $\geq$ 80\% (from 54\%)}'' and supplied a command containing \texttt{--fail-under=80} (R224). The target states the minimum acceptable coverage (E3.1) \blogs{B1, B21, B28},
while the command provides a way to check it (E3.2)~\blogs{B7, B8}. The documents show how to check the target, but do not show whether the check was run or passed. More generally, a verifiable specification identifies the scenarios that its checks must cover (E3.3)~\blogs{B5, B6} and the conditions for considering verification complete (E3.4)~\blogs{B27, B30}.

\textbf{P: Precise.}
To plan and carry out implementation work, an agent needs to understand both the conditions on which requirements depend and the scope it is expected to address. This category connects explicit prerequisites and constraints (\textit{Dependency Control}, P1)~\cite{ISO/IEC/IEEE-29148-2018}  with coverage of the intended requirements (\textit{Complete}, P2)~\cite{ISO/IEC/IEEE-29148-2018}.

\textit{P1. Dependency Control.}
An agent needs to know which assumptions and prerequisites must hold for a requirement to be realized. Declaring assumptions and dependencies (P1.1)\blogs{B17, B19} and stating constraints and scope boundaries (P1.2)\blogs{B30} make these conditions explicit. Linking them to individual requirements or tasks places that information alongside the work it affects (P1.3). For example, one repository has a statement ``\textit{Dependencies: Task 1 (sequential - needs root to link properly)}'' that maps to a task (R102). The entry identifies the prerequisite and explains why it must come first, providing a rationale for sequencing the work. Dependency verification and failure handling further specify how these conditions should be checked and what happens when they are not met (P1.4).

\textit{P2. Completeness.}
Completeness concerns whether the artifact set accounts for the requirements within the feature's stated scope, including functional and user requirements (P2.1)\blogs{B9, B13} and applicable exceptional behavior (P2.2)\blogs{B2, B5}. Scope exclusions identify what falls outside that boundary; missing or unresolved requirements within it indicate remaining gaps. For example, one repository included an edge-case table pairing ``\textit{Degenerate/near-zero triangles}'' with ``\textit{Filter or robust epsilon}'' (R164). The entry identifies a case the implementation needs to address, although the choice of handling remains open. Coverage also includes technical, operational, and quality constraints (P2.3)~\blogs{B15, B22}. Supporting plans record dependencies and how the work will be delivered (P2.4)~\blogs{B7}. Tracking unresolved specification items makes clear which parts of the feature still need decisions (P2.5)~\blogs{B6, B30}.

\textbf{I: Interpretable.}
An agent working across specifications, plans, and task descriptions needs to understand how the information fits together. Recognizable conventions help it locate requirements (\textit{Conforming}, I1)~\cite{ISO/IEC/IEEE-29148-2018}, compatible terminology and content preserve meaning across artifacts (\textit{Consistent}, I2)~\cite{ISO/IEC/IEEE-29148-2018}, and contextual guidance explains how the details relate to the intended product (\textit{Comprehensible}, I3)~\cite{ISO/IEC/IEEE-29148-2018}.

\textit{I1. Conformance.}
Conformance asks whether the artifacts follow the schema, template, or document conventions adopted for the project. Standardized structures establish where information belongs (I1.1)~\blogs{B7}, while stable, unique identifiers allow individual requirements to be referenced across documents (I1.2)~\repos{R385, R400}. Explicit template declarations identify the conventions the artifacts are expected to follow (I1.3)~\repos{R442, R446}. For example, one OpenSpec repository declared \texttt{schema: spec-driven} (R72). This gives a reader a named schema against which to check the artifacts.

\textit{I2. Consistency.}
As an agent moves between documents, the meaning of a requirement needs to remain the same. Terms and identifiers should refer to the same concepts, and descriptions of the same behavior under the same conditions should agree. Shared terminology and definitions establish these meanings (I2.1)~\repos{R72, R383}, while uniform notation keeps their expression consistent (I2.2)~\repos{R385, R441, R442}. In the same OpenSpec repository, seven requirements contained 35 \texttt{SHALL} and four \texttt{SHALL NOT} statements with no other modal variants. Each of its 21 scenarios contained one \texttt{GIVEN/WHEN/THEN} triple, illustrating consistency in normative wording and scenario structure.

Consistency also connects the specification to the planned implementation. One repository used the response fields \texttt{checks\_completed} and \texttt{checks\_skipped} identically in its specification, error cases, and implementation plan (R80). The matching names preserve the correspondence between the interface being specified and the one being planned (I2.3)~\blogs{B12, B16}. Repeating the same section organization across related artifacts provides another form of consistency (I2.4)~\blogs{B30}.

\textit{I3. Comprehensibility.}
The artifact set also needs to explain what is being built and how to navigate its description. Sections organized by purpose give readers a way into the details. For example, one repository organized its product requirements document into fifteen numbered sections, moving from the problem statement to non-goals (R166; I3.1)~\blogs{B30}. The same document introduced the product with a short statement: ``\textit{This is a prompt flow editor, not a chat app}.'' This tells the reader what the requirements describe and rules out a different interpretation of the product (I3.3)~\blogs{B3, B30}. When specifications, plans, and tasks occupy separate files, cross-document links and explanations of their relationships provide a route through the artifact set (I3.2)~\blogs{B23, B24}.

\textbf{C: Compartmentalized.}
An agent needs to distinguish what the product must do from the decisions and tasks involved in building it. This category connects a specification focused on required behavior (\textit{Scope Discipline}, C1)~\cite{ISO/IEC/IEEE-29148-2018} with supporting artifacts that guide implementation (\textit{Implementation-Planning Information}, C2)~\cite{kiro2025requirementsfirst,github2025speckit,fissionopenspec}.

\textit{C1. Scope Discipline.}
The product specification describes the capabilities users need and the constraints the product must satisfy. For example, one repository stated, ``\textit{As a network engineer, I want to query what the network looked like at a specific point in time}'' (R441). This describes the capability from the user's perspective (C1.1) \blogs{B2, B20}. In the analyzed examples, design decisions, technology stacks, and task lists appeared in separate plan and task artifacts (C1.2)~\blogs{B13, B18}. This gives implementation details a place while keeping the required behavior identifiable. Technical information can still belong in the product specification: a necessary compatibility, interface, or operating constraint is a requirement the implementation must satisfy.

\textit{C2. Implementation-Planning Information.}
Implementation-planning information concerns the guidance recorded in supporting plan and task artifacts. This includes the technical context in which the work will be implemented (C2.1)\repos{R100, R165}, what individual tasks are intended to produce (C2.2)\blogs{B9, B11}, and the data structures or interface contracts the implementation must respect (C2.4)\blogs{B10, B14}. For example, one repository supplied typed interfaces for stored entities, making the planned data structures and their types explicit (R100).

Traceability keeps these separate artifacts connected. \textit{Traceable Transition to Implementation} (C1.3)~\blogs{B3, B8} and \textit{Requirements-to-Implementation Traceability} (C2.3)~\repos{R441, R442} link requirements to the work intended to implement them. A task can then be read alongside the requirement it is meant to satisfy, keeping its purpose visible even when the implementation details appear in a different document.

\vspace{-3mm}
\subsection{RQ2. Evaluating the \EPIC Framework}

After developing \EPIC, we evaluated its practical relevance in two ways. First, we examined how practitioners interpreted \EPIC's practices. We then assessed whether repositories with higher specification-quality scores also showed better code quality and greater development activity.

\vspace{-2mm}

\subsubsection{Practitioner Assessment of \EPIC's Usefulness}
\label{sec:practitioner-results}

\begin{figure}
\centering
\includegraphics[width=0.9\linewidth]{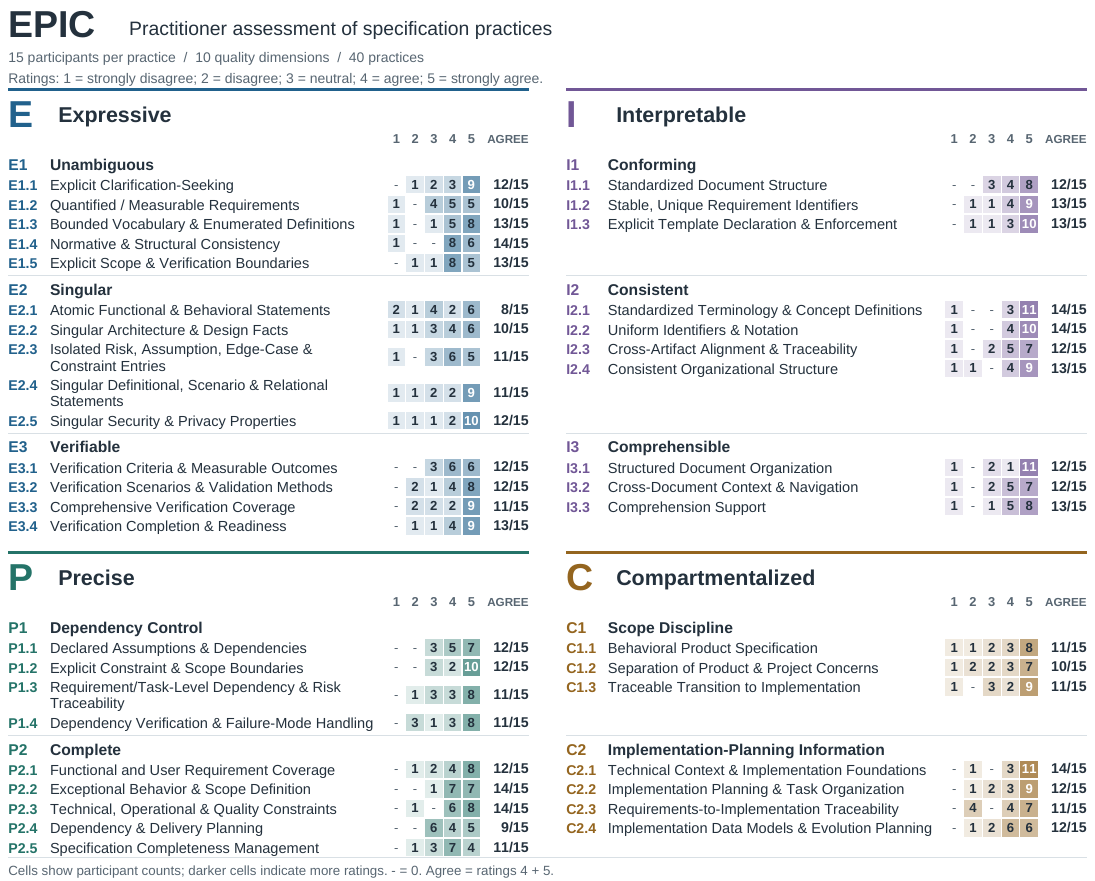}
\caption{The EPIC framework, organized into four categories, ten quality dimensions, and 40 practices, alongside evaluations from 15 practitioners. The columns to the right of each metric show the number of participants rating, from 1 (strongly disagree) to 5 (strongly agree).}
\label{fig:results_evaluation}
\vspace{-5mm}
\end{figure}

Overall, every metric in \EPIC's ten dimensions received agreement or strong agreement from a majority of participants as shown in Figure \ref{fig:results_evaluation}.

\textbf{Expressive.}
Participants endorsed the \textit{Expressive} practices for clarifying requirements (E1), keeping them singular (E2), and defining how they should be verified (E3) (Figure~\ref{fig:results_evaluation}).

\emph{Unambiguous (E1):}
Nine of 15 participants strongly agreed with \emph{explicit clarification-seeking} (E1.1). P2\footnote{Interview participants are identified as P1–P15; see \cite{suppl} for their demographics.} explained that when an agent faces an unresolved decision it may \pquote{just assume something}.
For decisions already made, 14 participants agreed with \emph{normative \& structural consistency} (E1.4). P4 explained that words such as \emph{should} or \emph{may} leave room for another interpretation: \pquote{the AI could say, okay, it should, so I'm going to go through the other way}. Their team therefore writes \pquote{you must do this, you must do that, so it has no open other interpretations}. Thirteen participants also agreed with \emph{explicit boundaries} (E1.5). P4 described how they kept the requested work within scope: \pquote{do not embrace the world. You have to focus on these tasks and only do this}.

\emph{Singular (E2):}
Eleven participants agreed with \emph{isolated risk \& edge-case entries} (E2.3). P2 explained that when each case has \pquote{its own bullet points \textup{[...]} the LLM tends to understand a little more}. Nine participants strongly agreed with \emph{singular definitions \& scenarios} (E2.4). P2 explained that keeping scenarios separate \pquote{really helps the agent to not be lost of what to do and how to do}. The connection between separation and checking also appeared in comments on \emph{singular security \& privacy properties} (E2.5), which received ten strongly agree ratings. P5 explained that recording these properties separately could allow the agent to check each one individually.

\emph{Verifiable (E3):}
Twelve participants agreed with \emph{measurable outcomes} (E3.1). P10 explained that \pquote{if [agent] doesn't know it's done, it's going to be impossible for it to stop the harness loop}. Twelve participants also agreed with \emph{verification methods} (E3.2);  \pquote{they say, yeah, 780 tests have passed. I don't really know anything about those tests and what they cover} (P3). Thirteen participants agreed with \emph{completion conditions} (E3.4). P2 described instances in which an agent stopped and announced that it had finished while leaving work incomplete. Stating completion conditions gave participants a basis for checking such claims and deciding whether the work could proceed. P9 explained that developers should always provide a \pquote{go signal or no go signal}.

Participants explained their lower ratings. For example, P3 explained that \pquote{now the AI figures this out} when discussing quantified requirements (E1.2), described agents as \pquote{trained to be really guarded about security and privacy} when discussing separate security and privacy properties (E2.5), and felt that splitting requirements could become \pquote{way too much to track}. However, \citet{pearce2022} found vulnerabilities in approximately 40\% of 1,689 programs generated across 89 security-relevant scenarios using an early version of Copilot.

\textbf{Precise.} A majority of participants agreed with every practice in this category, with five metrics receiving agreement from at least 12 of 15 participants (Figure~\ref{fig:results_evaluation}). 

\emph{Dependency Control (P1):} Ten participants strongly agreed with \emph{explicit constraints \& scope boundaries} (P1.2). P12 explained that explicitly stating constraints \pquote{prevents the design from going off the rails and requiring significant rework}, recalling cases where they had \pquote{ended up having to completely redo like big subsystems}. Eleven participants agreed with \emph{dependency verification \& failure-mode handling} (P1.4), including eight who strongly agreed. Their concern was that an implementation could assume its dependencies always work.  P4, whose system relied on external APIs, wanted it stated explicitly \pquote{what needs to be done when it fails, when it's out of service}.

\emph{Complete (P2):}
\emph{Exceptional behavior \& scope definition} (P2.2) and \emph{technical, operational \& quality constraints} (P2.3) each received agreement from 14 participants. P9 explained: \pquote{if you think there is enough, your customer will come and break the product in a way you've never seen}. Explicit exclusions also helped define the coverage needed: stating what falls outside the feature \pquote{helps a lot to funnel through what you actually need} (P6). For \emph{technical, operational \& quality constraints} (P2.3), P10 explained that different parts of a system may need different priorities. For example, \pquote{a billing system might have harder requirements around accuracy and replayability}, whereas a front-end component may have different needs. These expectations therefore need to be stated; otherwise, the agent is \pquote{making a decision even if I don't} (P10). P2.5 (\emph{specification completeness management}) was endorsed by 11 participants, who treated the record of decisions as context the agent can reuse. P11 kept a daily \pquote{log and make entries in a markdown, and then query against that with the LLM}, which helped \pquote{inject the context and history that I was looking for}.

Participants also explained their lower ratings in this category. Three participants disagreed with \emph{dependency verification \& failure-mode handling} (P1.4). P9 explained that \pquote{an agent will typically ask what you want to do when something fails}, suggesting that failure-handling decisions could be addressed when the problem arose. However, this expectation depends on the agent recognizing when it needs guidance and asking for it. Studies of coding agents show that detecting missing information and seeking clarification vary with the model, prompting, and agent configuration~\cite{vijayvargiya2026ambig,edwards2026ask}. For \emph{dependency \& delivery planning} (P2.4), six participants were neutral. P10 questioned the need for detailed planning early in a project, when \pquote{you just want to be able to run it right now}. This comment points to a question of timing: planning the delivery process may seem less necessary when the immediate goal is to get an initial implementation running.

\textbf{Interpretable.} This category received the highest proportion of agreement (Figure~\ref{fig:results_evaluation}). 


\emph{Conforming (I1):}
Thirteen participants agreed with \emph{stable, unique requirement identifiers} (I1.2). P5 used requirement numbers to trace product requirements into specifications and check that each had been addressed. P4's team also kept retired requirements visible so that a changed requirement \pquote{does not silently disappear}. Ten participants strongly agreed with \emph{explicit template declaration \& enforcement} (I1.3). P7 explained why declaring a convention alone may be insufficient: \pquote{The checks that you mechanize and make blockers on your process are the ones that actually get followed}.

\emph{Consistent (I2):}
Fourteen participants agreed with \emph{standardized terminology \& concept definitions} (I2.1), including 11 who strongly agreed. P2 described agents using different names for the same concept in different places. Consistency also mattered when information was distributed across documents. Twelve participants agreed with \emph{cross-artifact alignment \& traceability} (I2.3). P5 described a message schema maintained in one repository and its implementation in another. 


\emph{Comprehensible (I3):} Eleven participants strongly agreed with \emph{structured document organization} (I3.1). P10 explained that a clear structure helps both agents and developers locate relevant information: \pquote{both for humans to search, but also agents can go look at the table of contents and then go find the important piece}. Beyond finding information, participants wanted guidance on how to use it. Thirteen agreed with \emph{comprehension support} (I3.3). P2 explained that examples and guidance help \pquote{the human that are reading it, but also helps the agent to know how to perform}.

All seven ratings of 1 came from P3, who expressed confidence that agents could handle consistency: \pquote{in the AI world, it doesn't matter}. Research on trust in automation provides a possible explanation for such confidence. Users may initially consider automated systems trustworthy, but their confidence can decline after observing errors~\cite{dzindolet2003role,dietvorst2015algorithm}.

\textbf{Compartmentalized.} Participants described two needs: keeping the business problem visible (C1) and giving agents enough technical direction to implement it  (C2) (Figure~\ref{fig:results_evaluation}).

\emph{Scope Discipline (C1):} Eleven participants agreed with \emph{behavioral product specification} (C1.1). P5 explained: \pquote{When you mix those two things together, it becomes very hard to understand, like what is the actual business problem that you're trying to solve}. Participants also described where implementation guidance belonged. For \emph{separation of product \& project concerns} (C1.2), P2 placed that guidance in agent configuration, using \pquote{some skills and some rules that say how}.

\emph{Implementation-Planning Information (C2):} Fourteen participants agreed with \emph{technical context \& implementation foundations} (C2.1), including 11 who strongly agreed. P15 explained that without this context, an agent \pquote{can give the different technology solution for you}. Twelve participants agreed with \emph{implementation data models \& evolution planning} (C2.4). P2 described what happened when their team specified only the database: the generated entities \pquote{end up being like a direct copy of the database}, \pquote{because the agents tried to copy, because it's the reference it has}. Such a case illustrates how agents may use an available reference to fill gaps in the intended data model. Four participants disagreed with \emph{requirements-to-implementation traceability} (C2.3). P9 raised a concern about reviewing the volume of generated material: \pquote{nobody will be interested in reading all the things that agents created}. Automated checks could help address this concern. For example, P5 described using a custom agent skill to identify requirements that a codebase had not yet implemented.

\sisetup{
  detect-weight=true,
  mode=text,
  round-mode=places,
  round-precision=3
}

\begin{table*}[t]
\centering
\footnotesize
\setlength{\tabcolsep}{2.0pt}
\renewcommand{\arraystretch}{1.0}
\sisetup{detect-weight=true, mode=text}
\caption{Median code quality and activity by specification-quality band
($n=36/24/40$ for SonarQube; $40/26/43$ for GitHub).
Arrows: hypothesized directions; $\Delta$: relative top--bottom difference.
$p,\delta$: one-sided Mann--Whitney test and Cliff's effect size;
$\rho,p_{\rho}$: Spearman correlation and one-sided p-value.
H4 p-values are Benjamini--Hochberg-adjusted within each six-test family. Bold: $p<0.05$.}
\label{tab:hypotheses}


\begin{tabular}{@{}ll ccc c
S[table-format=1.4]
S[table-format=-1.3, retain-explicit-plus]
S[table-format=-1.3, retain-explicit-plus]
S[table-format=1.4]@{}}
\toprule
\multirow{2}{*}{\textbf{ID}}
& \multirow{2}{*}{\textbf{Measure}}
& \multicolumn{3}{c}{\textbf{Specification-quality band}}
& \multicolumn{3}{c}{\textbf{Top vs.\ bottom}}
& \multicolumn{2}{c}{\textbf{Spearman}} \\
\cmidrule(lr){3-5}\cmidrule(lr){6-8}\cmidrule(lr){9-10}
& & \textbf{Bottom} ($\leq 18$)
& \textbf{Middle} (19--22)
& \textbf{Top} ($\geq 23$)
& $\boldsymbol{\Delta}$
& \multicolumn{1}{c}{$\boldsymbol{p}$}
& \multicolumn{1}{c}{$\boldsymbol{\delta}$}
& \multicolumn{1}{c}{$\boldsymbol{\rho}$}
& \multicolumn{1}{c}{$\boldsymbol{p_{\rho}}$} \\
\midrule
H1 & Bugs/KLOC $\downarrow$
& 0.248 & 0.278 & 0.235 & $-5.1\%$
& 0.1007 & -0.171 & -0.183 & \bfseries 0.0345 \\
H2 & Vulnerabilities/KLOC $\downarrow$
& 0.178 & 0.274 & 0.092 & $-48.0\%$
& \bfseries 0.0161 & -0.286 & -0.134 & 0.0918 \\
H3 & Complexity/KLOC $\downarrow$
& 162.3 & 161.7 & 137.2 & $-15.4\%$
& \bfseries 0.0009 & -0.419 & -0.206 & \bfseries 0.0205 \\
\midrule
H4.A & Bug-fix commit rate $\downarrow$
& 20.4\% & 20.0\% & 11.8\% & $-42.4\%$
& \bfseries 0.0342 & -0.333 & -0.210 & \bfseries 0.0463 \\
H4.B & Pull requests $\uparrow$
& 89.0 & 210.0 & 202.5 & $+127.5\%$
& \bfseries 0.0359 & +0.241 & +0.170 & \bfseries 0.0463 \\
H4.C & PRs merged $\uparrow$
& 56.0 & 186.5 & 152.0 & $+171.4\%$
& \bfseries 0.0359 & +0.260 & +0.189 & \bfseries 0.0463 \\
H4.D & Issues $\uparrow$
& 22.0 & 107.5 & 58.0 & $+163.6\%$
& \bfseries 0.0364 & +0.229 & +0.161 & \bfseries 0.0475 \\
H4.E & Contributors $\uparrow$
& 3.0 & 5.5 & 12.5 & $+316.7\%$
& \bfseries 0.0359 & +0.241 & +0.201 & \bfseries 0.0463 \\
H4.F & Commits $\uparrow$
& 439.0 & 917.5 & 942.0 & $+114.6\%$
& \bfseries 0.0359 & +0.249 & +0.170 & \bfseries 0.0463 \\
\bottomrule
\end{tabular}
\vspace{-5mm}
\end{table*}

\subsubsection{Associations with Code Quality and Development Activity}

Table~\ref{tab:hypotheses} summarizes the associations between specification quality, code-quality indicators, and development activity. We observed that higher specification quality was associated with lower values on all three code-quality indicators, but the strength of the evidence varied.

\emph{H1. Reliability} was associated with specification quality across the full sample: repositories with higher scores tended to have fewer bugs per KLOC ($\rho = -0.183$, $p = 0.035$). However, the top and bottom bands had similar median bug densities (0.235 vs.\ 0.248), and their difference was not significant ($p = 0.101$), so H1 was supported only in the full-sample analysis. \emph{H2. Vulnerabilities} was lower in repositories with the highest specification scores. Top-band repositories had roughly half the median vulnerability density of bottom-band repositories (0.092 vs.\ 0.178 per KLOC; $p = 0.016$, $\delta = -0.286$). This difference did not extend to a significant correlation ($\rho = -0.134$, $p = 0.092$), so H2 was supported only in the top-versus-bottom comparison. \emph{H3. Structural maintainability} showed the most consistent association with specification quality. Top-band repositories had 15.4\% lower median cyclomatic complexity per KLOC (137.2 vs.\ 162.3; $p = 0.001$), with a medium effect size ($\delta = -0.419$). Complexity was also negatively correlated with specification score ($\rho = -0.206$, $p = 0.021$), supporting H3 in both analyses. \emph{H4. Repository activity} was higher in higher-score repositories. Top-band repositories had more pull requests, merged pull requests, issues, commits, and contributors, including a median of 12.5 contributors compared with 3 in the bottom band. All five activities were significant in both analyses after BH correction ($p_{\mathrm{BH}} \leq 0.048$). Bug-fix commits accounted for 11.8\% of scanned commits in the top band versus 20.4\% in the bottom band ($p_{\mathrm{BH}} = 0.034$), and their share decreased with specification score ($\rho = -0.210$, $p_{\mathrm{BH}} = 0.046$).

\vspace{-3mm}

\section{Discussion}
\label{sec:discussion}

\textbf{What \EPIC Adds to Requirements Quality.}
Requirements engineering has long established expectations for specifying what software must do and how its behavior should be verified~\cite{ISO/IEC/IEEE-29148-2018}. Our results show that these expectations still apply when the reader is an agent: seven of \EPIC's ten dimensions retain characteristics from ISO/IEC/IEEE~29148:2018, and none of the 30 grey-literature sources added a practice. What SDD changes is the unit of quality. The other three dimensions (P1, C1, C2) address a specification that comes with a plan and tasks, and practices such as cross-artifact alignment (I2.3) and requirements-to-implementation traceability (C2.3) concern the relationships among these documents rather than the wording of a single requirement. Participants explained why this matters when an agent does the implementation, describing agents filling gaps with assumptions, overlooking bundled requirements, and declaring incomplete work finished (Section~\ref{sec:practitioner-results}).


\textbf{Interpreting the Evaluation Results.}
Every \EPIC practice received agreement or strong agreement from a majority of participants, but support was not uniform. Normative wording (E1.4), exceptional behavior and technical constraints (P2.2, P2.3), standardized terminology (I2.1), and technical context (C2.1) each received agreement from 14 of 15 participants, which makes them a natural starting point for teams adopting \EPIC. Participants' reasons for lower ratings on other practices show that a practice's usefulness depends on context. \emph{Timing matters:} P10 questioned detailed delivery planning (P2.4) early in a project, when \pquote{you just want to be able to run it right now}. \emph{Reading cost matters:} P9 doubted that anyone would review agent-generated traceability (C2.3), since \pquote{nobody will be interested in reading all the things that agents created}. \emph{Ownership matters:} participants also differed on where implementation guidance belongs. P2, for example, placed it in agent configuration rather than in a separate plan (C1.2). Teams can therefore adopt \EPIC incrementally, starting with the broadly endorsed practices and adding others as a project grows and as the cost of an omission rises.

The repository associations in Table~\ref{tab:hypotheses} are observational, but participants' accounts suggest possible explanations for three of them: lower code complexity, the lower bug-fix rate, and the higher contributor activity. The first explanation concerns scope and constraints. Complexity had the most consistent association, with 15.4\% lower median complexity per KLOC in the top band and a decline across the full score range ($\rho = -0.206$). Participants described agents expanding a task beyond what was asked, choosing an unintended technology, or copying the only data model available, and missing constraints forcing the redesign of whole subsystems (Section~\ref{sec:practitioner-results}). Explicit boundaries (E1.5), constraints (P1.2), technical context (C2.1), and data models (C2.4) leave fewer of these decisions to the agent.

The second explanation is verifiability. Participants explained that an agent cannot reliably judge whether its own work is correct or finished: agents may report completion they have not reached, and developers may be unable to interpret the tests an agent writes (Section~\ref{sec:practitioner-results}). Stating measurable outcomes (E3.1), verification methods (E3.2), and completion conditions (E3.4) gives both the agent and the developer a way to check the work without relying on the agent's own report. At least 12 of 15 participants endorsed each of these practices. When a specification states how the work will be checked, defects may be caught before they need later fixes. This explanation is consistent with the lower bug-fix share in top-band repositories (11.8\% of scanned commits vs.\ 20.4\% in the bottom band), and the share also declined across the full score range ($\rho = -0.210$).

The third explanation is that the specification may also serve as onboarding documentation. Unclear documentation and poorly documented project structure are known barriers to contribution~\cite{steinmacher2014hard}, and repository documentation has been linked to contribution activity~\cite{gaughan2025introduction}. In an SDD project, the specification states the scope, constraints, and completion conditions that both a new contributor and a new agent session need. This role may help explain why top-band repositories had a median of 12.5 contributors compared with 3 in the bottom band, and more than twice as many pull requests (202.5 vs.\ 89). The relationship may also run the other way: projects with more contributors may invest more in their specifications. These explanations need further testing.

\textbf{Applying \EPIC in Development Workflows.}
\EPIC provides quality criteria that can be applied wherever specification and planning artifacts guide implementation, including workflows without a dedicated SDD framework. SDD frameworks such as OpenSpec and GitHub Spec Kit could use \EPIC to guide the clarification questions they ask and the revisions they suggest~\cite{fissionopenspec,github2025speckit}, and teams could check specifications in CI/CD workflows whenever specifications or plans change, building on automated requirements-smell detection~\cite{femmer2017rapid}.

\begin{wrapfigure}{r}{0.5\textwidth}
  \centering
  \vspace{-3mm}
  \includegraphics[width=0.5\textwidth]{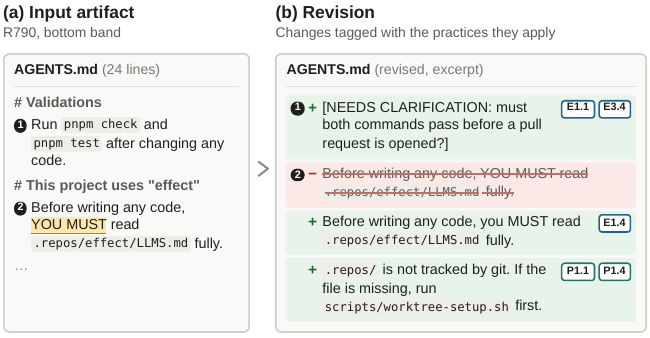}
  \caption{Applying the \EPIC skill: (a) original instructions from R790; (b) instructions revised by the skill, with numbered markers linking problems to changes tagged with \EPIC practices.}
  \label{fig:skill-example}
    \vspace{-3mm}
\end{wrapfigure}

\textit{Packaging \EPIC as an agent skill.}
A more portable approach is to give the framework directly to the coding agent. As an initial implementation, we packaged \EPIC as an agent skill~\cite{agentskills}: a \texttt{SKILL.md} instruction file with supporting references. The skill can be reused across projects, which answers Gao et al.'s call to consolidate domain knowledge that each project otherwise writes on its own~\cite{gao2026registry}.

The skill encodes \EPIC's four categories, ten dimensions, and 40 practices, along with an adapted scoring rubric, criteria for what counts as evidence, and examples from the analyzed repositories. In review mode, it examines specification and planning artifacts together, scores each dimension from 0 to 3, and cites passages from the artifacts to support each score. In authoring mode, which covers both drafting new artifacts and revising existing ones, it applies the practices and marks unresolved decisions for clarification. In both modes, it separates practices that do not apply to a feature from information that is missing, and it does not penalize the former.

Figure~\ref{fig:skill-example} illustrates the authoring mode. In authoring mode, the skill found that R790's instructions required a file that only a setup step creates, and the instructions never mention that step. The skill added the step, based on the repository, and turned the policies that the file leaves open into clarification questions. These examples show how the skill works; they do not measure how reliably it scores or revises specifications. Because the associations in RQ2 are correlational, a higher score also does not guarantee better code. The skill, its reference files, and the complete outputs for both examples are available in our replication package~\cite{suppl}.

\textbf{How Much Guidance Will Agents Need?}
Some participants rated practices lower because they expected agents to handle the issue themselves; P3, for example, expected agents to keep artifacts consistent on their own (Section~\ref{sec:practitioner-results}). Agents may handle some omissions well, but whether an agent detects missing information and asks for it varies with the model, prompting, and agent configuration~\cite{vijayvargiya2026ambig,edwards2026ask}. A specification also has a second reader: the developer who checks the agent's work, who, as participants described, may be unable to judge claims of completion without stated outcomes and verification criteria. Future benchmarks could use \EPIC to compare specifications that include, weaken, or omit a practice, such as dependency failure handling (P1.4), and measure across models and tasks whether the agent asks, decides acceptably, or violates a constraint. Such evidence would show which practices agents can compensate for and which decisions still require the developer.

\vspace{-1mm}

\section{Limitations}
\label{sec:threats}
\textbf{Limitations in developing \EPIC.} Like all empirical studies, our study has limitations. We acknowledge that we may have overlooked quality dimensions or specification practices relevant to SDD when developing \EPIC. To reduce this risk, we grounded the quality lens in ISO/IEC/IEEE~29148:2018 and SDD tool guidance, derived practices from repository artifacts, and examined practitioner recommendations and interview feedback. These perspectives helped us identify and refine relevant practices.

During the analysis of repository artifacts, scoring and exemplar selection involved researcher judgment. To mitigate potential bias, four researchers first developed the rubric, and we conducted two calibration rounds to assess its consistent application. We complemented the repository findings with grey-literature guidance.

\textbf{Limitations in evaluating \EPIC.} 
The evaluation of \EPIC provides an initial empirical foundation for SDD practices. However, we did not invite participants to apply \EPIC in their own projects. Future work will evaluate \EPIC in use with developers of different backgrounds and experience levels and reassess its coverage and applicability as SDD workflows and agent capabilities evolve.

We chose semi-structured interviews to evaluate \EPIC because they allowed us to explore participants' explanations, practical constraints, and examples in detail. The 15 participants had varied backgrounds and levels of software development experience, and we screened them for SDD experience to obtain feedback grounded in practice. To reduce participant response bias~\cite{dell2012yours}, we presented \EPIC as a draft, welcomed criticism, and used neutral follow-up questions. The disagreements and explanations reported alongside agreement show that participants were willing to challenge the framework.

The limited repository sample constrains estimate precision and our ability to detect smaller effects. We report effect sizes alongside p-values and apply BH correction~\cite{benjamini1995controlling}. Without multivariate controls or stratified analyses, project characteristics and development practices remain potential confounders, so the associations cannot establish causation. SonarQube results also vary across languages and project structures, while our focus on public, English-language repositories limits generalizability. We treat these findings as exploratory, drawing on complementary statistical tests and practitioner evaluations to examine \EPIC from multiple perspectives.

\vspace{-1mm}
\section{Conclusion}
\label{sec:conclusion}




We introduced \EPIC, an empirically grounded framework comprising four categories, Expressive, Precise, Interpretable, and Compartmentalized, ten quality dimensions, and 40 practices that guide the writing and review of specification and planning artifacts in SDD. Our future work follows two directions. First, we will evaluate \EPIC in real development settings with developers of different experience levels, including longitudinal studies in student capstone projects, to understand how the framework affects specification practices and the resulting software. Second, an urgent research priority is to develop benchmarks that examine how coding agents interpret and act on this guidance.  Ultimately, as P12 reflected, \pquote{the spec cannot take responsibility for the future; I have to take responsibility for it. To take responsibility for something, I must understand it.}

\section*{Data Availability}
The research artifacts for this study are publicly available on the companion website \cite{suppl}, and the \EPIC framework is available at \cite{epic_website_2026}. 



\bibliographystyle{config/ACM-Reference-Format}
\bibliography{references}

\end{document}